\documentclass{article}
\usepackage{spconf,amsmath,graphicx}
\usepackage{bm}
\usepackage{multirow}
\usepackage{booktabs}
\usepackage{cases}

\usepackage{cite}
\usepackage{hyperref}

\usepackage{stfloats}
\usepackage{algorithm} 
\usepackage{algorithmic} 
\usepackage{multirow} 
\usepackage{amsmath}
\usepackage{xcolor}
\usepackage{booktabs}   
\usepackage{threeparttable}
\usepackage{amsfonts}
\usepackage{url}
\usepackage{caption}
\usepackage{array}
\usepackage{makecell}
\usepackage{subcaption}
\usepackage{textcomp}  

\title{Self-Refining Deep Symmetry Enhanced Network for Rain Removal}

\name{Hong Liu\textsuperscript{1}~~~~Hanrong Ye\textsuperscript{1*}~~~~Xia Li\textsuperscript{1}~~~~Wei Shi\textsuperscript{1}~~~~Mengyuan Liu\textsuperscript{2}~~~~Qianru Sun\textsuperscript{3}\thanks{\scriptsize *Corresponding author. 
This work is supported by National Natural Science Foundation of China (NSFC, No. U1613209), Scientific Research Project of Shenzhen City (JCYJ20170306164738129), Shenzhen Key Laboratory for Intelligent Multimedia and Virtual Reality (No.ZDSYS201703031405467). This project is partly by NExT++ research supported by National Research Foundation, Singapore under its IRC@SG Funding Initiative. Code will be available at \url{https://github.com/prismformore/SDSEN}}}

\address{\textsuperscript{1}Key Laboratory of Machine Perception, Shenzhen Graduate School, Peking University \\
	\textsuperscript{2} School of EEE, Nanyang Technological University \\
	\textsuperscript{3} School of Computing, National University of Singapore \\
	\{hongliu, leoyhr, ethanlee, pkusw\}@pku.edu.cn, nkliuyifang@gmail.com, qianrusun@comp.nus.edu.sg}

\begin{document}
%
\maketitle
%
\begin{abstract}
	Rain removal aims to remove the rain streaks on rain images. The state-of-the-art  methods are mostly based on Convolutional Neural Network~(CNN). However, as CNN is not equivariant to object rotation, these methods are unsuitable for dealing with the tilted rain streaks. 
	To tackle this problem, we propose Deep Symmetry Enhanced Network~(DSEN) that is able to explicitly extract the rotation equivariant features from rain images.
	In addition, we design a self-refining mechanism to remove the accumulated rain streaks in a coarse-to-fine manner. This mechanism reuses DSEN with a novel information link which passes the gradient flow to the higher stages. Extensive experiments on both synthetic and real-world rain images show that our self-refining DSEN yields the top performance.	

\end{abstract}
\begin{keywords}
Image Restoration, Rotation Equivariance, CNN


\end{keywords}

\vspace{-0.6em}
\section{Introduction}
\label{sec:intro}
\vspace{-0.6em}

Rain degenerates the visibility of images and affects vision algorithms severely. By restoring the background information in rain scenes, rain removal is helpful for improving many computer vision systems including autonomous driving, intelligent photography and surveillance camera. On this task, convolutional neural network~(CNN) based methods have manifested promising performance. 
DetailNet~\cite{fu2017_cvpr} focuses on the high-frequency detail and regresses negative residual information with an end-to-end pipeline.  
JORDER~\cite{jorder}  removes rain streaks with a deep recurrent dilated network. DID-MDN~\cite{zhang2018cvpr_density} proposes a multi-stream CNN utilizing density information of rain.

\begin{figure}[t]
\centering
\vspace{-0.5em}
\includegraphics[scale=0.19]{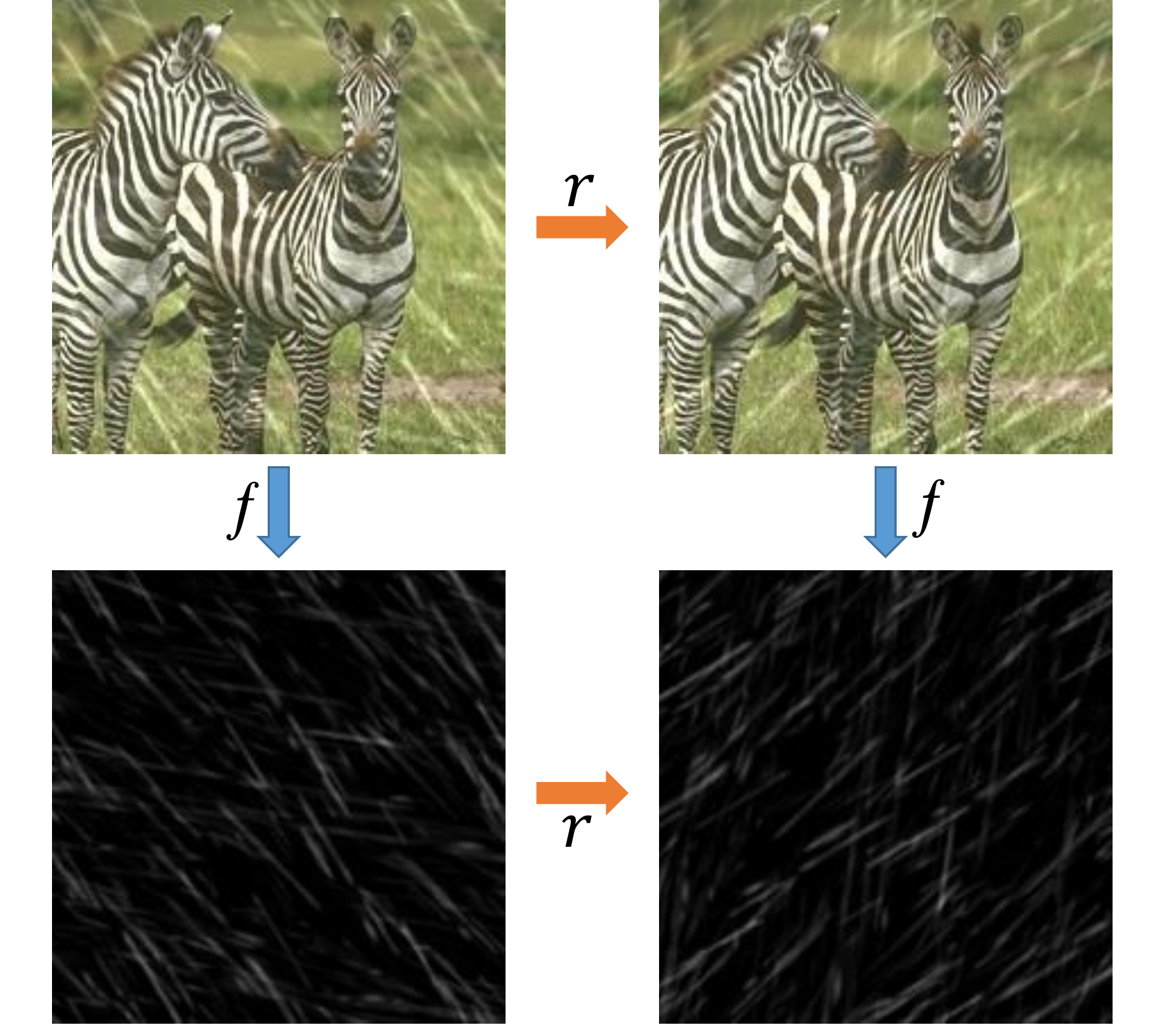}
\vspace{-0.5em}
\caption{
	 Equivariance: Extracted rain layer is expected to rotate according to the rotation of rain streak. $r$ denotes a 90\textdegree~clockwise rotation on rain streak and $f$ is the mapping of a rain extraction model.}

\vspace{-1.7em}
\label{img_equi}
\end{figure}

As the rain streaks are always tilted at various angles, we expect the generated rain layer to follow the rotation of rain streaks. This property is called ``rotation equivariance'' as shown in Fig.~\ref{img_equi}. However, the conventional convolution operation in CNNs does not possess rotation equivariance, which decreases the ability of CNNs to learn representation for rotated objects~\cite{weiler_recnn,marcos_revfn} like tilted rain streaks. Therefore, existing CNN based methods have difficulty in processing tilted rain streaks by their nature.

To tackle this problem, we enhance CNN with rotation equivariance and propose a symmetry enhanced CNN called Deep Symmetry Enhanced Network~(DSEN).
Our ideas are as follows. First, we introduce a special convolution layer, which is mathematically designed with rotation equivariance property. This special convolution layer is stacked to extract \textit{symmetry enhanced features} of rain, which have stronger representation ability for tilted rain streaks than regular features.  Then, as symmetry enhanced features need to be aggregated before decoding rain layer, we propose an aggregation block, which avoids the information loss in down-sampling of other methods~\cite{weiler_recnn}.
Finally, we use a regular convolution layer to decode the rain layer for image restoration.

As rain streaks overlap with each other, the \textit{accumulation effect} makes it difficult to remove all rain streaks in a single stage. To solve this problem, we design a multi-stage framework called self-refining mechanism based on the traits of image restoration task. To utilize information of the previous stage effectively, this framework is designed with novel information links between every two adjacent stages. 

In summary, our contributions are three-fold:
(1) The first symmetry enhanced CNN for rain removal, which can handle the tilted rain streaks in images explicitly.  To the best of our knowledge, our method is the first symmetry enhanced CNN that achieves the state-of-the-art performance on pixel-level image processing tasks.
(2) A new symmetry aggregation block which collects symmetry enhanced features without losing the orientation information. 
(3)  A nontrivial self-refining mechanism is proposed to remove rain streaks in a coarse-to-fine manner. 




\begin{figure*}  
	\vspace{-1.3em}
	\centering
	\includegraphics[scale=0.27]{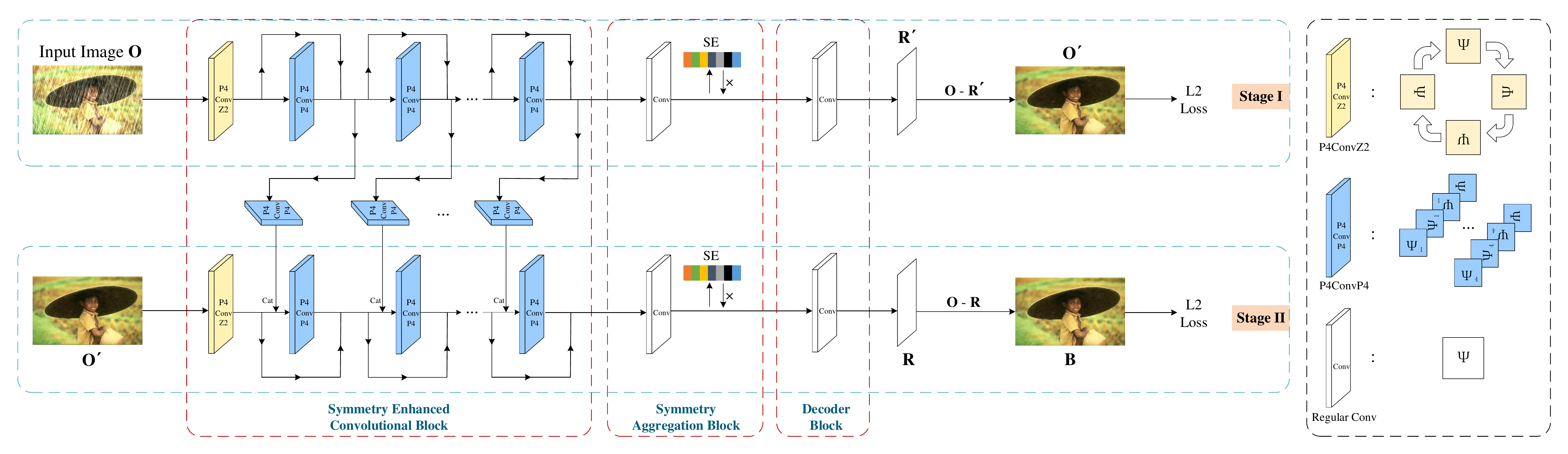}
	\vspace{-1em}
	\caption{A two-stage example of Self-refining DSEN. 
	The output of stage  I (\textbf{O\'}) is reused as the input of stage II. 
	P4ConvZ2~(yellow) contains one filter with four rotations~(corresponding to 90\textdegree, 180\textdegree, 270\textdegree~and 360\textdegree). P4ConvP4~(blue) contains four filters with four rotations. Best viewed in color.}
	\vspace{-1.6em}
	\label{img_sdsen}
\end{figure*}

\vspace{-0.6em}
\section{Method}


\vspace{-0.6em}
\subsection{Deep Symmetry Enhanced Network}
\vspace{-0.6em}
As shown in Fig.~\ref{img_sdsen}, DSEN consists of three function blocks: Symmetry enhanced convolutional block, symmetry aggregation block and decoder block. To process images with arbitrary sizes, the proposed model is designed as a fully convolutional network with neither fully connected layer nor pooling operation. 

\begin{figure*}[!h]
	\vspace{-1.1em}
	\centering
	
	\begin{subfigure}{0.17\textwidth}
		\includegraphics[width=\textwidth]{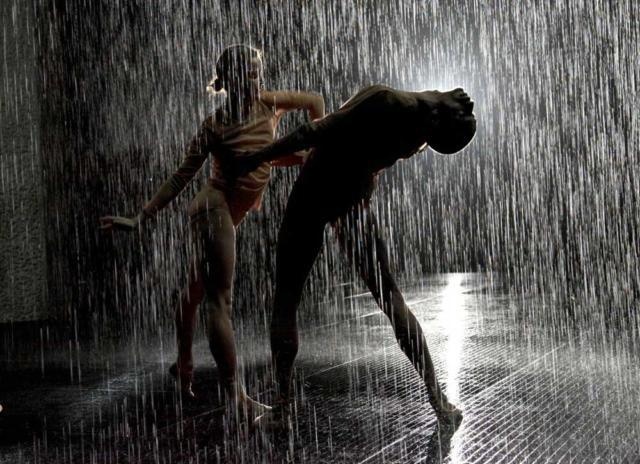}
	\end{subfigure}
	\begin{subfigure}{0.17\textwidth}
		\includegraphics[width=\textwidth]{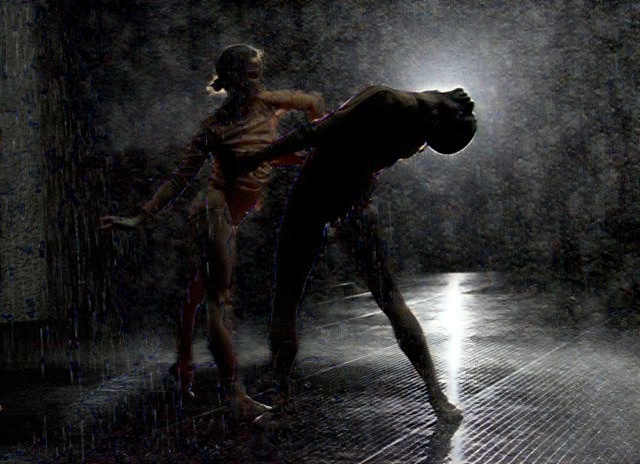}
	\end{subfigure}
	\begin{subfigure}{0.17\textwidth}
		\includegraphics[width=\textwidth]{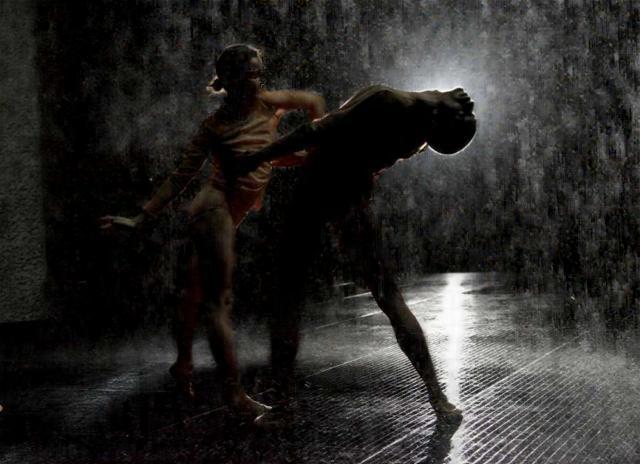}
	\end{subfigure}
	\begin{subfigure}{0.17\textwidth}
		\includegraphics[width=\textwidth]{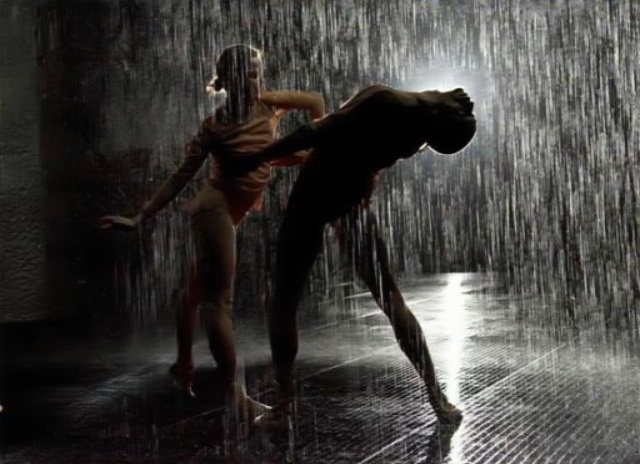}
	\end{subfigure} 
	\begin{subfigure}{0.17\textwidth}
		\includegraphics[width=\textwidth]{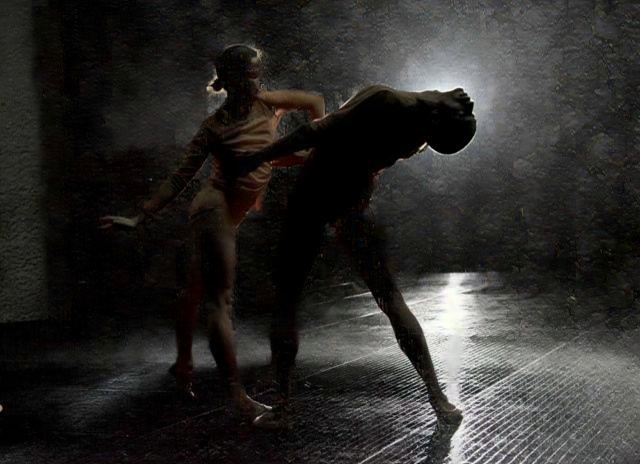}
	\end{subfigure}
	
	\begin{subfigure}{0.17\textwidth}
		\includegraphics[width=\textwidth]{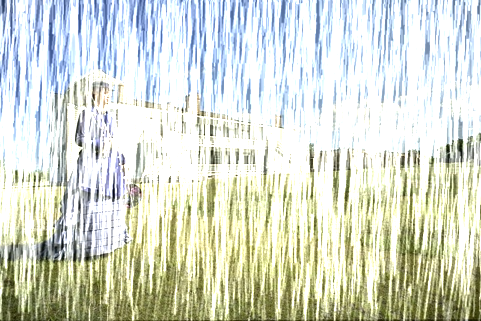}
		\caption{Input \protect\\ 6.99 / 0.2369}
	\end{subfigure}
	\begin{subfigure}{0.17\textwidth}
		\includegraphics[width=\textwidth]{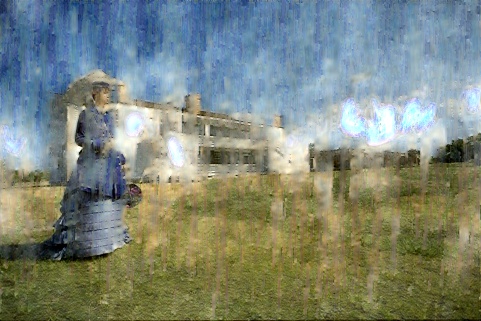}
		\caption{DetailNet~\cite{fu2017_cvpr}\protect\\ 17.91 / 0.5422}
	\end{subfigure}
	\begin{subfigure}{0.17\textwidth}
		\includegraphics[width=\textwidth]{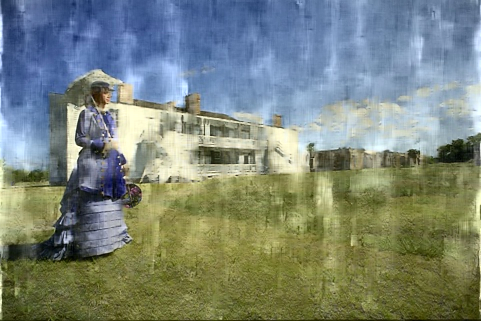}
		\caption{JORDER-R~\cite{jorder}\protect\\ 18.13 / 0.6122}
	\end{subfigure}
	\begin{subfigure}{0.17\textwidth}
		\includegraphics[width=\textwidth]{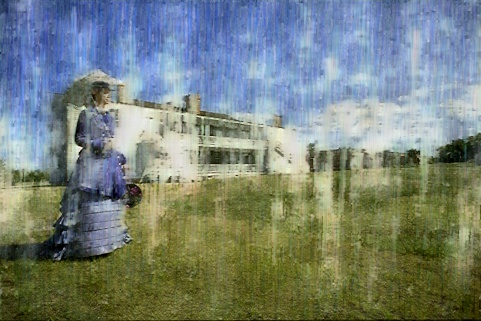}
		\caption{DID-MDN~\cite{zhang2018cvpr_density}\protect\\ 17.60 / 0.5163}
	\end{subfigure}
	\begin{subfigure}{0.17\textwidth}
		\includegraphics[width=\textwidth]{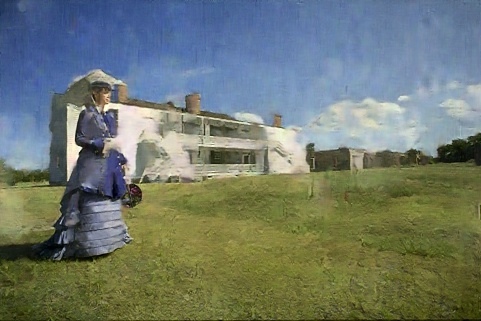}
		\caption{S-DSEN~(ours) \protect\\ \textbf{22.37} / \textbf{0.7237}}
	\end{subfigure}

	\vspace{-1em}
	\caption{Restored samples  of real-world and synthetic images. Our method generates images with fewer rain streaks and more details evidently. PSNR / SSIM values are provided for synthetic cases.}
	\vspace{-2em}	
	\label{img_sub}
	
\end{figure*}

\textbf{Symmetry Enhanced Convolutional Block}: 
A lot of effort has been made to extend the equivariance of neural networks~\cite{le2010_tiledcnn,gens2014_dsn,dieleman2015,dieleman2016,marcos_revfn,worrall2017_harmonic,cohen2018_spherical,zhang2017_access,esteves2018_ptn}. 
Tiled CNNs~\cite{le2010_tiledcnn} change the weight sharing strategy to a tiled pattern for learning rotation symmetry of data. 
Deep Symmetry Networks~\cite{gens2014_dsn} generalize CNN and compute features over arbitrary symmetry groups.  Dieleman et al.~\cite{dieleman2015,dieleman2016} show that, by rotation of feature maps, symmetry enhanced CNNs are able to learn rotationally equivariant representations and achieve success on tasks including galaxy morphology prediction, plankton classification and binary segmentation~(pixel-level classification). Cohen and Welling~\cite{gcnn} propose a mathematical framework for the study of rotation equivariance based on group theory, and design a rotationally equivariant convolution called G-convolution.
Instead of rotating feature maps, G-convolution rotates filters to achieve rotation equivariance. The general definition~\cite{gcnn} of G-convolution from  group $H$ to group $I$~($H \to I$) is:
\vspace{-0.6em}
\begin{equation}
\hat{G}_If^H(i) = \sum_{k=1}^{K^{l-1}} \sum_{h \in H}  f_k(h)(\hat{T}_i \psi_k(h)),
\label{gcnn}
\vspace{-0.8em}
\end{equation}
\noindent where $\hat{G}_I$ denotes the G-convolution operator of group $I$, $\hat{T}_i$ is the operator of transformation $i$ from group $I$~($i \in I$) and $h$ is a transformation of group $H$. Both filter $\psi_k$ and feature maps $f^H$ are functions defined on group $H$. When Eq.~\ref{gcnn} satisfies $I=p4$, we call it a $p4$ group equivariant convolution~($p4$ G-convolution). 

In this block, we adopt $p4$ G-convolution as basic operation to keep equivariant to rotation transformation while extracting high-level features through deep architecture. $p4$ G-convolution composes of two types: $p4$ G-Convolution on $\mathbb{Z}^2$~(P4ConvZ2) and $p4$ G-convolution on $p4$~(P4ConvP4).  To process the plane information of input images, P4ConvZ2 is compulsory as the first layer. For the rest layers, P4ConvP4 is stacked to increase the network capacity for abstracting high level features, and the depth can be adjusted depending on practical need. Residual connection~\cite{resnet} is adopted on each P4ConvP4 layer. 

For $p4$ G-convolution~($\hat{G}_{p4}$), since every transformation $i \in p4$ can be decomposed into a translation $x$ and a rotation $r$ from the corresponding point group~\cite{gcnn} $C_4$~(rotation by 90\textdegree, 180\textdegree, 270\textdegree~and 360\textdegree), and for the same reason $h \in p4$ can be decomposed into a translation $y$ and rotation $s$, P4ConvP4 can be rewritten as:
\vspace{-0.2em}
\begin{equation}
\hat{G}_{p4} f^{p4}(x, r) =
\sum_{k}  \sum_{y \in \mathbb{Z}^2}  \sum_{s \in C_4}   f_k^{p4}(y,s)  (\hat{T}_x (\hat{T}_r \psi_k(y,s))),
\label{fml_p4cp4}
\end{equation}
\noindent where $\hat{T}_x$ denotes the operator of translation $x$ and $\hat{T}_r$ denotes the operator of rotation $r$. Since group $\mathbb{Z}^2$ contains only translation transformation~($y$), P4ConvZ2 is written as:
\vspace{-0.2em}
\begin{equation}
\hat{G}_{p4} f^{\mathbb{Z}^2}(x, r) = \sum_{k} \sum_{y \in \mathbb{Z}^2}  f_k^{\mathbb{Z}^2}(y)(\hat{T}_x (\hat{T}_r \psi_k(y))).
\label{fml_p4cz2}
\end{equation}
\vspace{-0.2em}
As shown in Fig.~\ref{img_sdsen} and detailed in \cite{gcnn}, P4ConvZ2 layer contains one filter with four rotations. Its input $f^{\mathbb{Z}^2}$ and filter $\psi_k$ are both defined on plane $\mathbb{Z}^2$, while the output $\hat{G}_{p4} f^{\mathbb{Z}^2}$ is a function on the group $p4$. P4ConvP4 layer contains four filters with four rotations. Its input, filters and output are all defined on group $p4$. For simplicity, we stack only 4 P4ConvP4 layers after one P4ConvZ2 layer in this block. There are 10 regular channels in all filters  with the kernel size of $5\times 5$.

\textbf{Symmetry Aggregation Block}: The symmetry enhanced feature maps~(called \textit{G-feature maps}) generated by symmetry enhanced convolutional block have 4 orientation channels~(corresponding to four rotations in $C_4$) for each regular channel. To aggregate rotationally equivariant features, orientation pooling method~\cite{weiler_recnn} pools over the orientation channels within each regular channel on binary segmentation task. However, since different orientation channels encode rain streaks from corresponding aspects, and rain streaks are not strictly perpendicular to each other, orientation pooling method causes information loss. To remedy this problem, we propose to apply a  $5\times 5$ convolution on all orientation channels of G-feature maps. Additionally, as the channels of aggregated feature maps contribute differently to rain layer, an  Squeeze-and-Extraction~\cite{senet} block is adopted to compute attention value of each channel explicitly. 

\textbf{Decoder block}:  This module is a convolution layer with $1 \times 1$ kernel size to decode and generate the three-channel~(RGB) rain layer \textbf{R}. 
As rain image can be decomposed into the background layer and the rain layer, we obtain the restored background \textbf{B} by subtracting rain layer \textbf{R} from original image \textbf{O}~\cite{jorder,lixia}. The restored image is supervised by the per-pixel Euclidean loss~(L2 loss) function.

\vspace{-0.6em}
\subsection{Self-refining Mechanism}
\vspace{-0.6em}
For image restoration problem, the output of system shares the same domain as its input. For example, output of the rain removal model is still an image of the same scene with fewer rain streaks. Thus the output of image restoration networks can be reused as input for finer processing. This multi-stage recurrent pipeline  encourages the reuse of parameters to exploit the potential capacity of base models. This pipeline could be used to tackle the accumulation effect of rain streaks overlapping. However, simply repeating the network faces gradient vanishing problem at training, which leads to bad performance as shown in the following experiment section. Therefore, we need to improve the gradient flow between deeper and shallower stages for the recurrent structure.

To this end, we design a novel self-refining mechanism with a stage-wise link called \textit{skip concatenation}, which resembles a combination of dense connection~\cite{densenet} and skip connection~\cite{resnet}. We illustrate our framework with a two-stage example of DSEN with self-refining mechanism~(S-DSEN) in Fig.~\ref{img_sdsen}. The pipeline of stage I follows DSEN. In stage II, the G-feature maps of each P4ConvP4 layer at stage I is extracted and processed by an intermediate P4ConvP4 convolution layer generating the intermediate G-feature maps. The intermediate G-feature maps are concatenated with input of the same layer at stage II and then fed into the next layer. Different from dense connection, the links only exist between adjacent stages with stage-wise weight sharing, thus the increase of stage number would not add any model parameters while increasing refining ability. We set the maximum recurrent stage number to 8 in experiment.  

\vspace{-0.6em}
\section{Experiments and Discussion}
\vspace{-0.6em}

\vspace{-0.6em}
\subsection{Benchmark}
\vspace{-0.6em}

\textbf{Datasets} 
\textit{Rain100H}~\cite{jorder} is synthesized with five rain streak directions. Zhang et al.~\cite{zhang2017_arxiv} synthesize dataset with 800 image pairs containing rain streaks of various intensities and orientations, hence we call it \textit{Zhang800}. 
According to the official split, in \textit{Rain100H} there are 1800 image pairs for training and 100 for testing; In \textit{Zhang800}, there are 700 image pairs for training and 100 for testing. We randomly select 100 image pairs from training set for validation on both datasets.


\noindent \textbf{Implementation Details}
Since rain removal is a local image processing problem, to accelerate the training process, training images are randomly cropped to the size of $64\times 64$. Deep learning models are trained on an Nvidia GTX 1080Ti GPU with Pytorch implementation, setting the batch size to 64 unless otherwise stated. LeakyReLU is used as activation function with negative slope set to 0.2. We adopt Adam algorithm~\cite{adam} for optimization and set the learning rate to 0.0005 initially, which is divided by 10 at step 15000 and 17500. All trainable parameters are randomly initialized. Two commonly used image quality measures are adopted: Structural Similarity Index~(SSIM)~\cite{image_quality} and Peak Signal to Noise Ratio~(PSNR)~\cite{psnr}, for quantitative study. 


\vspace{-0.6em}
\subsection{Overall Analysis}
\vspace{-0.6em}
In this part, we study the overall effectiveness of DSEN. 
We design a CNN counterpart with exactly the same structure as DSEN but replace all the G-convolutions by regular convolutions and double the number of channels per filter to keep parameter size approximately invariant~\cite{gcnn}. Since data augmentation helps CNNs approximate equivariance, for comparison, we apply random rotation~(-30\textdegree~$\sim$ 30\textdegree) and C4 data augmentation~(random rotation by 90\textdegree, 180\textdegree and 270\textdegree) on training images of regular CNN~(CNN+DA and CNN+C4DA).

Experiment results are reported in Table \ref{tab_q1}, manifesting that DSEN outperforms its regular CNN counterpart on all indicators assuredly with fewer model parameters. Remarkably, data augmentation strategy only helps increase one indicator while decreasing all others. According to our survey, training images rotation is seldom adopted in rain removal. We speculate that image rotation changes the data distribution and thus is inappropriate for this problem. Consequently, symmetry enhanced CNNs are the better solution for label symmetry learning in rain removal.

\begin{table}[t]
	\centering
	\vspace{-2em}
	\caption{PSNR and SSIM of regular CNN and DSEN on \textit{Rain100H} and \textit{Zhang800}. The best results are highlighted in bold. }
	\vspace{-1em}
	\scalebox{0.7}{
		{
			\begin{tabular}{ccccccccccccc}
				\toprule
				Dataset & \multicolumn{2}{c}{\textit{Rain100H}} & \multicolumn{2}{c}{\textit{Zhang800}} &
				\\
				\midrule
				Metric & PSNR &SSIM  & PSNR &SSIM  &{Parameter Size}
				\\
				\midrule
				CNN & 23.36 & 0.7557 & 22.88 & 0.8127 &52113
				\\
				CNN+DA & 22.21 & 0.7055 & \textbf{24.04} & 0.8044 & 52113
				\\
				CNN+C4DA & 22.74 & 0.7320 & 22.49 & 0.7891 & 52113
				\\
				DSEN & \textbf{23.60} & \textbf{0.7618} & 23.39 & \textbf{0.8162} & 50958
				\\
				\bottomrule
			\end{tabular}}}
			\label{tab_q1}
		\vspace{-0.9em}
		\end{table}

\begin{table}
	\centering
	\caption{PSNR and SSIM on \textit{Rain100H} and \textit{Zhang800}. The best results are highlighted in bold.}
	\vspace{-1em}
	\scalebox{0.7}{
		{
			\begin{tabular}{ccccccccccccc}
				\toprule
				Dataset & \multicolumn{2}{c}{\textit{Rain100H}} & \multicolumn{2}{c}{\textit{Zhang800}}
				\\
				\midrule
				Metric & PSNR &SSIM  & PSNR &SSIM & &
				\\
				\midrule
				DSEN\underline{~~}maxpool & 22.83 & 0.7359 & 22.30 & 0.7916
				\\
				DSEN\underline{~~}avgpool & 22.86 & 0.7382 & 22.30 & 0.7916
				\\
				DSEN\underline{~~}w/o\underline{~~}SE & 23.46 & 0.7595 & 22.53 & 0.8035
				\\
				DSEN & 23.60 & 0.7618 & 23.39 & 0.8162
				\\
				\midrule
				MCNN & 23.70 & 0.7226  & 22.90 & 0.7965
				\\
				MCNN+SR  & 26.36 & 0.8395  &  23.52 & 0.8362
				\\
				MDSEN & 23.98 & 0.7295 & 23.48 & 0.8000
				\\
				MDSEN+SC  & 25.84 & 0.8182 & \textbf{23.79} & 0.8269
				\\
				S-DSEN  & \textbf{27.16} & \textbf{0.8589} & 23.64 & \textbf{0.8379}
				\\
				\bottomrule
			\end{tabular}}}
			\label{tab_q2}
			\vspace{-1.5em}
		\end{table}

\vspace{-0.6em}
\subsection{Component Analysis}
\vspace{-0.6em}

In this experiment, we study the effect of  model components, especially symmetry aggregation block and self-refining mechanism.

\textbf{Symmetry Aggregation Block} We compare the proposed symmetry aggregation block with orientation pooling methods~\cite{weiler_recnn} including orientation max-pooling~(DSEN\underline{~~}maxpool) and orientation average-pooling~(DSEN\underline{~~}avgpool). For ablation study of the Squeeze-and-Excitation~(SE) block, we remove it from DSEN (DSEN\underline{~~}w/o\underline{~~}SE). 
Experiment results in Table~\ref{tab_q2} indicate that DSEN outperforms its variants on all indicators.

\textbf{Self-refining Mechanism~(SR) }To validate SR, which works on not only DSEN but also regular CNNs, we construct two multi-stage networks based on the CNN in Q1: with or without SR~(MCNN+SR and MCNN). Both networks adopt 8-stage recurrent architecture. Experiment results in Table~\ref{tab_q2} show that with self-refining mechanism, the performance of regular CNN soars with  SSIM and PSNR improved by 16.18\% and 11.22\% separately  on \textit{Rain100H}. Similarly, eight-stage S-DSEN raises SSIM and PSNR by 17.74\% and 13.26\% on \textit{Rain100H} compared with its counterpart without SR~(MDSEN). 
The performance of plain recurrent networks~(MCNN and MDSEN) are even worse than their single-stage base models in Table~\ref{tab_q1} concerning SSIM. This serves as evidence for the importance of stage-wise information links in recurrent architecture and shows the validity of self-refining mechanism. What's more, we design an alternative stage-wise skip connection~(SC), which is similar to the residual connection of ResNet~\cite{resnet}, in place of SR for comparison. SC does not concatenate the intermediate feature maps with the next-stage input of the same layer, but simply adds on it. 

In order to investigate how the performance varies with the increase of stages, we report the experiment results of S-DSEN with several different stage number in Table~\ref{tab_q2_sr}. Compared with the single-stage DSEN in Table~\ref{tab_q2}, SSIM values of 2-stage S-DSEN\underline{~~}s2,  4-stage S-DSEN\underline{~~}s4 and 6-stage S-DSEN\underline{~~}s6 increase by 5.38\%, 9.32\% and 11.84\% separately on \textit{Rain100H} dataset. 

\begin{table}
	\centering
	\vspace{-2em}
	\caption{Comparison of S-DSEN with different stage number on \textit{Rain100H} and \textit{Zhang800}.}
	\vspace{-1em}
	\scalebox{0.7}{
		{
			\begin{tabular}{ccccccccccccc}
				\toprule
				Dataset & \multicolumn{2}{c}{\textit{Rain100H}} & \multicolumn{2}{c}{\textit{Zhang800}} &
				\\
				\midrule
				Metric & PSNR &SSIM  & PSNR &SSIM  
				\\
				\midrule
				S-DSEN\underline{~~}s2  & 25.04 & 0.8028 & 23.53 & 0.8270 
				\\
				S-DSEN\underline{~~}s4  & 26.24 & 0.8328 & 23.74 & 0.8340 
				\\
				S-DSEN\underline{~~}s6  & 26.90 & 0.8520 & 23.66 & 0.8364 
				\\
				\bottomrule
			\end{tabular}}}
			\label{tab_q2_sr}
			\vspace{-1em}
		\end{table}

\vspace{-0.6em}
\subsection{Comparison with the State-of-the-art Methods}
In this subsection, we compare our S-DSEN with other rain removal methods in Table~\ref{tab_q3}. As our approach is the first symmetry enhanced CNN used in image restoration, we compare it with regular CNN based methods. DSEN and S-DSEN achieve the state-of-the-art performance.

As subjective evaluation is important for rain removal algorithms, a comparison of restored images generated by several methods from both real-world and synthetic rain images are shown in Fig.~\ref{img_sub}.  The proposed method generates images with fewer rain streaks and more details.


\begin{table}
	\centering
	\caption{Comparison of quantitative results. The best are highlighted in bold and the second best are underlined.}
	\vspace{-1em}
	\scalebox{0.7}{
	{
	\begin{tabular}{ccccccccccccc}
		\toprule
		Dataset & \multicolumn{2}{c}{\textit{Rain100H}} & \multicolumn{2}{c}{\textit{Zhang800}}
		\\
		\midrule
		Metric & PSNR &SSIM  & PSNR &SSIM 
		\\
		\midrule
		ID~\cite{kang2012_tip} & 14.02 & 0.5239  & 18.88 &0.5832 
		\\
		DSC~\cite{luo2015_iccv} & 15.66 & 0.4225 & 18.56 & 0.5996 
		\\
		LP~\cite{li2016_cvpr_prior} & 14.26 & 0.5444 & 20.46 & 0.7297
		\\
		DetailNet~\cite{fu2017_cvpr}  & 22.26  & 0.6928 & 21.16 &  0.7320
		\\
		JORDER~\cite{jorder}  & 22.15 & 0.6736 & 22.24 & 0.7763
		\\
		JORDER-R~\cite{jorder}  & 23.45 & 0.7490 & 22.29 & 0.7922 
		\\
		DID-MDN~\cite{zhang2018cvpr_density} & \underline{24.53} & \underline{0.7990} & 21.71 & 0.7810
		
		\\
		\midrule
		DSEN~(ours)  & 23.60 & 0.7618	& \underline{23.39} & \underline{0.8162}
		\\
		S-DSEN~(ours)  & \textbf{27.16} & \textbf{0.8589} & \textbf{23.64} & \textbf{0.8379}
		\\
		\bottomrule
	\end{tabular}}}
	\label{tab_q3}
	\vspace{-1.5em}
\end{table}

\vspace{-0.6em}
\section{Conclusion}
\vspace{-0.6em}
In this paper, we proposes a novel symmetry enhanced network to explicitly remove the tilted rain streaks from rain images. Moreover, a handy self-refining mechanism, which works on both regular CNNs and symmetry enhanced CNNs, is designed to tackle the accumulation effect of rain streaks.
For future work, we plan to leverage the idea of explicit structure modeling~\cite{Ma_NIPS2017, Ma_2018_CVPR, Sun_2018_CVPR, Sun_2018_ECCV} to improve our method for fixing pixel-level flaws and generating high-quality images.

\vspace{-0.6em}

\bibliographystyle{IEEEbib}
\bibliography{refs}

\end{document}